\documentclass[twocolumn,twoside]{article}
\usepackage{graphics}
\newcommand{\h}{\linebreak \hspace*{3ex}}
\newcommand{\hb}{\\ \hspace*{2ex}}

\begin{document}
\title{STELLAR OBJECTS OF EXTRAGALACTIC ORIGIN IN THE GALACTIC HALO}
\author{V.A.\,Marsakov$^{1}$, T.V.\,Borkova$^{1}$\\[2mm] 
\begin{tabular}{l}
 $^1$ Southern Federal University\hb
 Rostov-on-Don 344090 Russia, \\
 {\em marsakov@ip.rsu.ru}, {\em borkova@ip.rsu.ru}\\
\end{tabular}
}
\date{}
\maketitle

ABSTRACT.
We identified globular clusters and field stars of extragalactic 
origin and investigated their chemical, physical, and kinematical 
properties. This objects as supposed was captured by the
Galaxy at different times from debris of the dwarf satellite
galaxies disrupted by its tidal forces. The results are follows. 
(1) The majorities of metal-poor stellar objects in the Galaxy 
have an extragalactic origin. (2) The masses of the accreted 
globular clusters decrease with the removal from the center and 
the plane of the Galaxy. (3) The relative abundances of chemical 
elements in the accreted and genetically connected stars are 
essentially distinguished. (4) The accreted field stars demonstrate 
the decrease of the relative magnesium abundanses with an 
increase in sizes and inclinations of their orbits. (5) The 
stars of the Centaurus moving group were born from the matter, 
in which star formation rate was considerably lower than in 
the early Galaxy. On the base of these properties was made a 
conclusion that with the decrease of the masses of the dwarf 
galaxies in them simultaneously decrease the average masses 
of globular clusters and the maximum masses of supernova SNe\,II. 
Namely latter fact leads to the decrease of the relative 
abundances of $\alpha$-elements in their metal-poor stars.
\\[1mm]

{\bf Galaxy (Milky Way), stellar chemical composition, 
accreted stellar objects, halo, Galactic evolution.}\\[2mm]


Still very recently they assumed  that our Galaxy was 
formed from the united proto-galactic cloud, and all its 
objects are genetically connected together. However, the 
numerous observations of the last years demonstrate to 
us compelling evidence that the Galaxy closely interacts 
with the less massive satellite galaxies and gradually 
destroying them, captured their interstellar matter, 
separate stars and globular clusters. In particular, we 
are currently observing the disruption of dwarf galaxy 
Sagittarius by tidal forces from the Galaxy. About ten 
globular clusters are associated with this galaxy. The 
massive globular cluster M\,54 is generally believed to 
be the nucleus of the system. The galactic orbital 
elements of same else clusters also suggest that they 
were captured from various satellite galaxies. There 
are convincing proofs that even $\omega$\,Cen, the largest 
known globular cluster of the Galaxy, which is close 
to the Galactic center and has retrograde orbit, was 
the nucleus of a dwarf galaxy in the past. The theory 
of dynamical evolution predicts the inevitable 
dissipation of clusters through the combined actions of 
two-body relaxation, tidal destruction, and collisional 
interactions with the Galactic disk and bulge. Indeed, 
traces of the tidal 
interaction with the Galaxy in the shape of extended 
deformations (tidal tails) have been found in all the 
clusters for which high-quality optical images were 
obtained. It is even established for $\omega$\,Cen that, 
after the last passage 
through the plane of the disk, this cluster lost slightly 
less than one percent of its mass in the form of stars. 
Thus, even in the nearest solar neighborhood, we may 
attempt to identify stars of extragalactic origin. It is 
interesting to investigate the distinctive properties of 
stellar objects of extragalactic origin and to estimate 
their relative number. 

It turned out that metal rich ($[Fe/H] > -1.0$) objects 
form the rapidly revolving and completely flattened 
subsystem of the thick disk. But metal-poor objects are 
divided into two types of populations also. Is relied that 
the metal-poor stars of field with the peculiar velocities 
are less than the critical value and globulars with the 
extremely blue horizontal branches form the genetically 
connected with the thick disk spherical, slowly rotating 
subsystem of their own halo with the insignificant, but 
the different from zero radial and vertical metallicity 
gradients. The high velocity field stars and globulars 
with the horizontal branches of intermediate color form 
the spherical subsystem of external accreted halo, 
approximately into two and one-half of times of larger 
size than two previous. In this case the absence in it 
of the metallicity gradients, the predominantly elongated 
orbits, the large number of stars with retrograde galactic 
rotation, and often small ages confirm hypothesis about 
their extragalactic origin.

\begin{figure}[t!]
\resizebox{\hsize}{!}
{\includegraphics{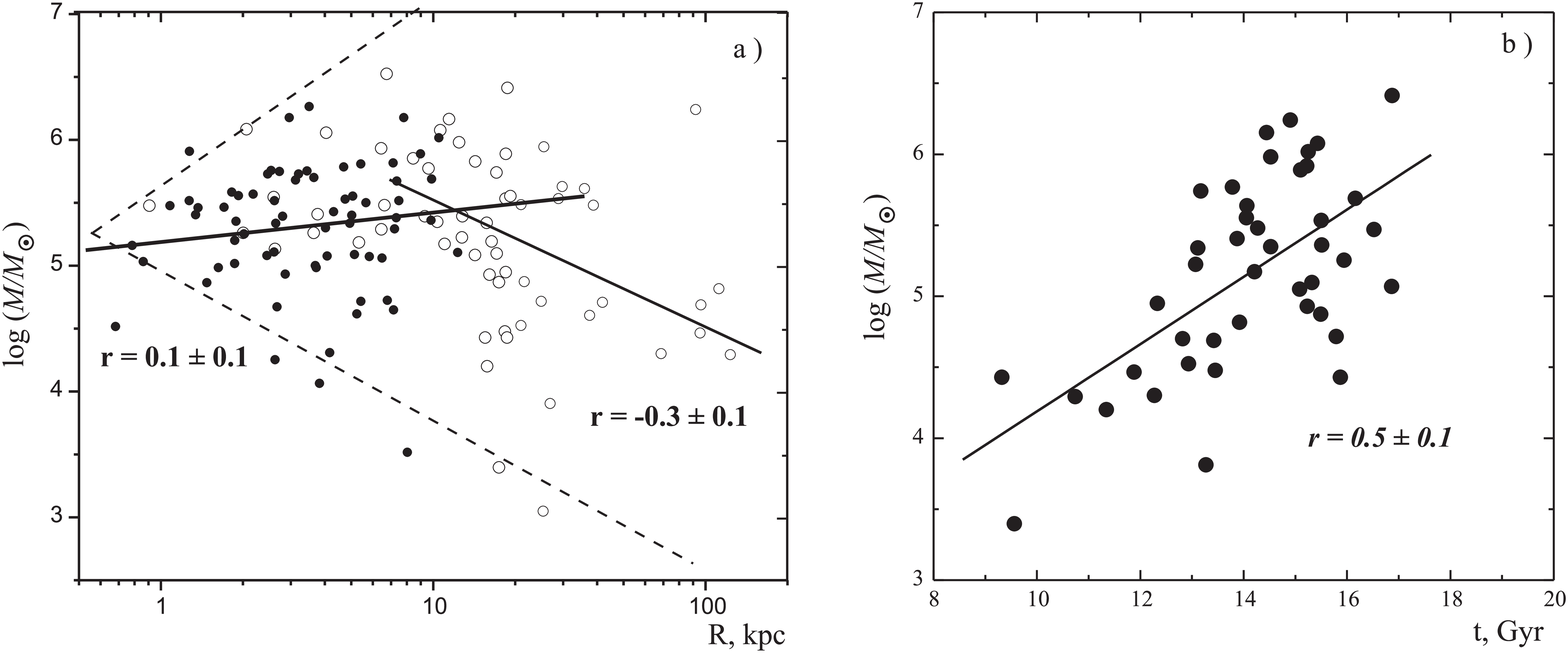}}
\caption{The relationships between the mass and the observed 
galactocentric distance (a), and between the mass and the age (b). 
The solid circles denote the genetically connected globular 
clusters, and open circles~-- the accreted ones. The solid 
lines are least square fits for genetically connected and 
accreted globulars. The dotted line in diagram (a) restrict 
the region of slow evolution of the globular clasters. The 
corresponding correlation coefficients are indicated. It is 
seen the good correlation for accreted clusters in both diagrams.}
\label{fig_1}
\end{figure}

Very important for understanding of nature of accreted globulars 
is one of their properties. They demonstrate the dependence of 
mass on the galactocentric distance (Borkova, Masakov,~2000). 
Solid lines in the diagram $R_a$~--  mass (Fig.\,\ref{fig_1}\,a) 
are the regression straight lines for genetically connected and 
accreted globulars. As we see almost all accreted clusters are 
into the region of the slow evolution of globular clusters the 
existence of which is theoretically grounded. Therefore they 
did not undergo the significant action of dissipation and 
dynamic friction; i.\,e., in external halo their initial mass 
distribution was preserved almost without the change. It is 
evident that the genetically connected clusters do not 
reveal a change in the average mass with an increase in the 
distance from the Galactic center. While for the accreted 
clusters the observed anticorrelation is different from 
zero far beyond ranges of errors. Simultaneously accreted 
clusters demonstrate the decrease of average mass, also, 
with the decrease of age (Fig.\,\ref{fig_1}\,b). The genetically 
connected globulars of this effect do not reveal. It seems 
that the globulars with small masses frequently are formed 
beyond the limits of the Galaxy. Moreover, the greater the 
dimensions of their present galactic orbit, the less their 
mass and the ages in average.  Hence the conclusion: the 
globular clusters with anomalously small mass and age are 
formed predominantly in the such low massive satellite 
galaxy, which even being located at sufficiently great 
distances from the Galactic center, lose their globulars 
under the action of its tidal forces.

It is unlikely that the interstellar matter from 
which the stars of own and accreted halo were 
formed has experienced an exactly coincident chemical 
evolution. Therefore, it would be interesting to search 
for subtle differences between them that could shed light 
on the histories of star formation inside 
and outside the single proto-galactic cloud.
Owing to the position of the Sun in the Galactic
plane, we have an opportunity to observe the stars of
all its subsystems in the immediate vicinity of the Sun
and to analyze in detail their chemical composition.

According to current conception the evolution time of close binary stars 
that subsequently explode as SNe\,Ia is short, $\approx 1$~Gyr. 
Exclusively higher-mass ($M>8 M_\odot$) stars exploding as 
type~II supernovae (SNe\,II) are currently believed to have 
enriched the interstellar medium with heavy elements at 
earlier stages. Their characteristic evolution time is only 
$\approx 30$~Myr. Almost all of the nuclei of $\alpha$-elements 
are formed in SNe\,II while the bulk of iron-peak
elements is ejected into the interstellar space during SN\,Ia explosions. 
Calculations show that the yield of the 
$\alpha$-elements depends strongly on the stellar mass. 
Therefore, the relative 
abundances of  $\alpha$-elements ([$\alpha$/Fe]) 
in the ejecta of SNe\,II with different mass can 
differ markedly. Hence, the variations in the upper boundary 
of the initial 
mass function for stars that exploded 
inside and outside the Galaxy can be estimated from 
the relative abundances of various elements in genetically 
related and accreted stars. Concurrently because of the 
difference between 
the evolution times of SNe\,II and SNe\,I we can try to trace 
the star formation rate for this stellar ensemble by the 
coordinates of the characteristic knee in its [$\alpha$/Fe]--[Fe/H]
diagram toward the sharp decrease in the relative abundance 
of the $\alpha$-elements with increasing total 
heavy-element abundance at the onset of SNe\,Ia explosions, i.\,e., 
$\sim 1$~Gyr later.

The best-studied $\alpha$-element is 
magnesium because they exhibit several absorption lines in 
the visible spectral range. For the analysis, we took data 
from our compiled catalog of 
spectroscopically determined magnesium abundances (Borkova 
and Marsakov~2005). Almost all of the magnesium abundances 
in nearest stars determined 
by synthetic modeling of high-dispersion spectra and published
before January~2004 were gathered in this catalog. 
The relative magnesium abundances in the catalog were derived 
from 1412~spectroscopic determinations in 31~publications for 
867~dwarfs and subgiants using a three-pass iterative averaging 
procedure with a weight assigned to each primary source and each
individual determination. The internal accuracy of the 
relative magnesium abundances for metal-poor 
([Fe/H]$<-1.0$) stars is $\varepsilon$[Mg/Fe]=$\pm 0.07$. 

We justified the choice of 
the peculiar stellar velocity relative to the local 
standard of rest $V_{res}=175$~km\,s$^{-1}$ as a criterion 
for separating the nearest
thick-disk and halo stars. In identifying the stars of an 
extragalactic origin (which were 
called here accreted stars), we assumed that the stars born in 
a monotonically collapsing single proto-galactic cloud could 
not be in retrograde orbits. We included all of the stars with 
high orbit energy, i.\,e. of high peculiar velocities 
$V_{res} >240$~km\,s$^{-1}$, as have all stars with retrograde 
orbits, in the group of presumably accreted stars.

\begin{figure}
\resizebox{\hsize}{!}
{\includegraphics{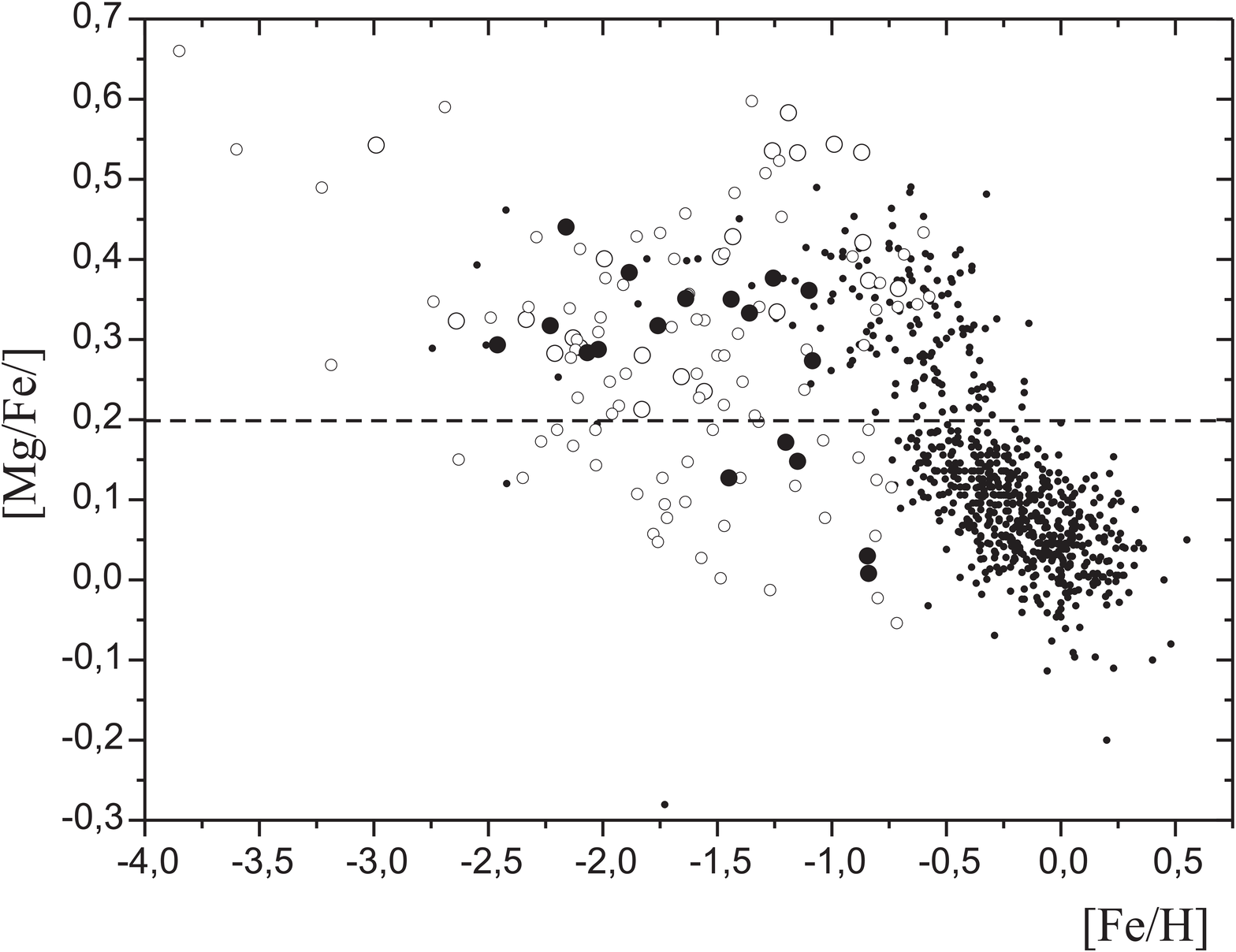}}
\caption{Metallicity vs. relative magnesium abundance 
for all of the stars in the catalog. The crosses, asterisks, 
and circles indicate thin- and thick-disk stars, own halo 
stars, and presumably accreted stars. The filled 
circles highlight the members of the Centaurus moving group 
among the accreted stars. The dashed line was drawn through 
$[Mg/Fe] = 0.2$.}
\label{fig_2}
\end{figure}

Figure\,\ref{fig_2} shows the metallicity~-- relative magnesium 
abundance diagram for our catalog. It is seen that
the accreted objects subsequently formed the bulk of the Galactic
halo. (By this term we mean all of the objects that were born outside 
the single proto-galactic cloud, i.\,e., in the nearest satellite 
galaxies or in isolated proto-galactic fragments, and that subsequently 
escaped from them under the Galactic tidal forces.) We see also
from the Figure\,\ref{fig_2} that the relative magnesium abundances 
in the own-halo 
stars are virtually independent of metallicity and that all 
stars of own halo lie above the dashed line drawn through 
$[Mg/Fe]=0.2$. This behavior of own-halo stars suggests that, 
at least in the initial 
stage of its formation the interstellar matter in the early Galaxy 
either was well mixed or SNe\,II of the same mass exploded in all 
local volumes. In contrast, the presumably accreted stars exhibit a 
large spread in relative magnesium abundances in Fig.\,\ref{fig_2} that extends
to negative [Mg/Fe]. The anomalously low relative magnesium abundances 
in some of the accreted stars are usually explained by an extremely 
low star formation rate in the dwarf satellite galaxies where these
stars were born. However our analysis of the 
relative magnesium and europium abundances in a small sample of nearby 
field stars showed that large portion of the presumably accreted 
stars exhibited 
an [Eu/Mg] ratio that differed sharply from its Galactic value (Borkova 
and Marsakov~2004). Since the relative yield of these elements depends 
solely on the masses of the SN\,II progenitor stars where they are 
synthesized, we believe that a more likely mechanism of the magnesium 
abundance variations in accreted stars is the difference between the 
initial mass functions in their parent dwarf satellite galaxies.
Therefore, it is interesting to try to identify genetically related 
stars in the accreted halo.

We identified from our catalog of the stars of moving group, 
which was supposedly lost by the dwarf galaxy, whose center as 
supposed was the cluster $\omega$\,Cen. In the [Mg/Fe]--[Fe/H] 
diagram (see Fig.\,\ref{fig_2}), all of them  lie along a narrow strip. 
This behavior 
resembles the expected [Mg/Fe]--[Fe/H] relation derived in a 
closed model of chemical evolution, which is independent evidences 
for the genetic relationship between the identified stars. Hence, 
the low relative magnesium abundances in the 
metal-richest stars of this group resulted from the SN\,Ia explosions 
that began in their parent proto-galactic cloud and that ejected a 
large number of iron atoms into the interstellar medium and reduced the 
[Mg/Fe] ratio. The considerably lower metallicity of the knee point 
in this diagram than that in the Galaxy suggests that the stars of 
the Centaurus moving group were formed from matter in which the 
star formation rate was considerably lower than that in the early 
Galaxy. The high initial relations [Mg/Fe] evidences that, at 
least in this, presumably initially massive 
($M\approx 10^9 M_\odot $)  disrupted satellite galaxy 
(Tshuchiya, et al.,~2003) the mean masses of the SN\,II 
progenitor stars were the same as those in our Galaxy.
It is known that according to numerical 
simulations of dynamical processes during the interaction
of galaxies (Abadi et al.~2003) the satellite galaxies are disrupted 
and lose their stars only after dynamical friction 
reduces significantly the sizes of their orbits and drags them 
into the Galactic plane. Less massive satellite galaxies are 
disrupted even before their orbits change appreciably under 
tidal forces. Therefore lost by them stars as a rule must be in higher 
and more distant orbit. Let us verify this theoretical assumption.

\begin{figure}
\resizebox{\hsize}{!}
{\includegraphics{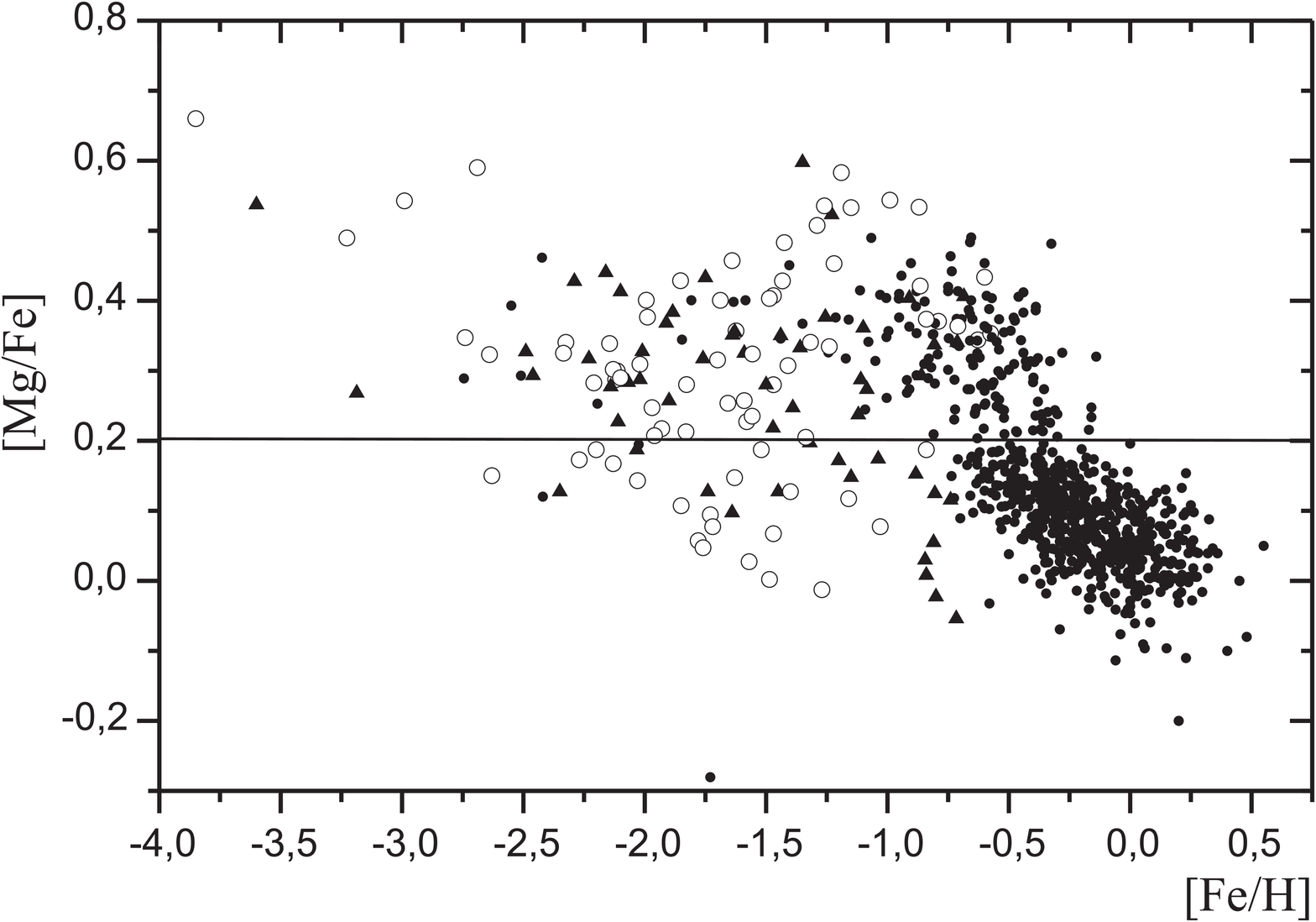}}
\caption{Relative magnesium abundances vs. metallicity. The 
crosses and circles indicate the genetically related stars 
and presumably accreted stars.  The filled circles represent 
presumably accreted stars with azimuthal velocities in the 
ranges $\Theta > 50$~km\,s$^{-1}$.}
\label{fig_3}
\end{figure}

From Fig.\,\ref{fig_3} it is evident that only slowly rotating around 
the Galactic center stars with the small relations $[Mg/Fe]< 0.2$ are 
observed with $[Fe/H]>-1.0$. (Centaurus moving group also have the 
angular momentum close to zero with retrograde rotation.) 
Consequently we may to assume that all slow stars were born in 
the sufficiently massive satellite galaxies. Moreover the star 
formation rate in them was actually lowered, in comparison with 
the Galaxy, since the stars of them demonstrate less metal rich 
"knee point". While the overwhelming majority 
magnesium-poor and simultaneously metal-poor accreted stars 
fell within the range $|\Theta |<50$~km\,s$^{-1}$.

\begin{figure}[t!]
\resizebox{\hsize}{!}
{\includegraphics{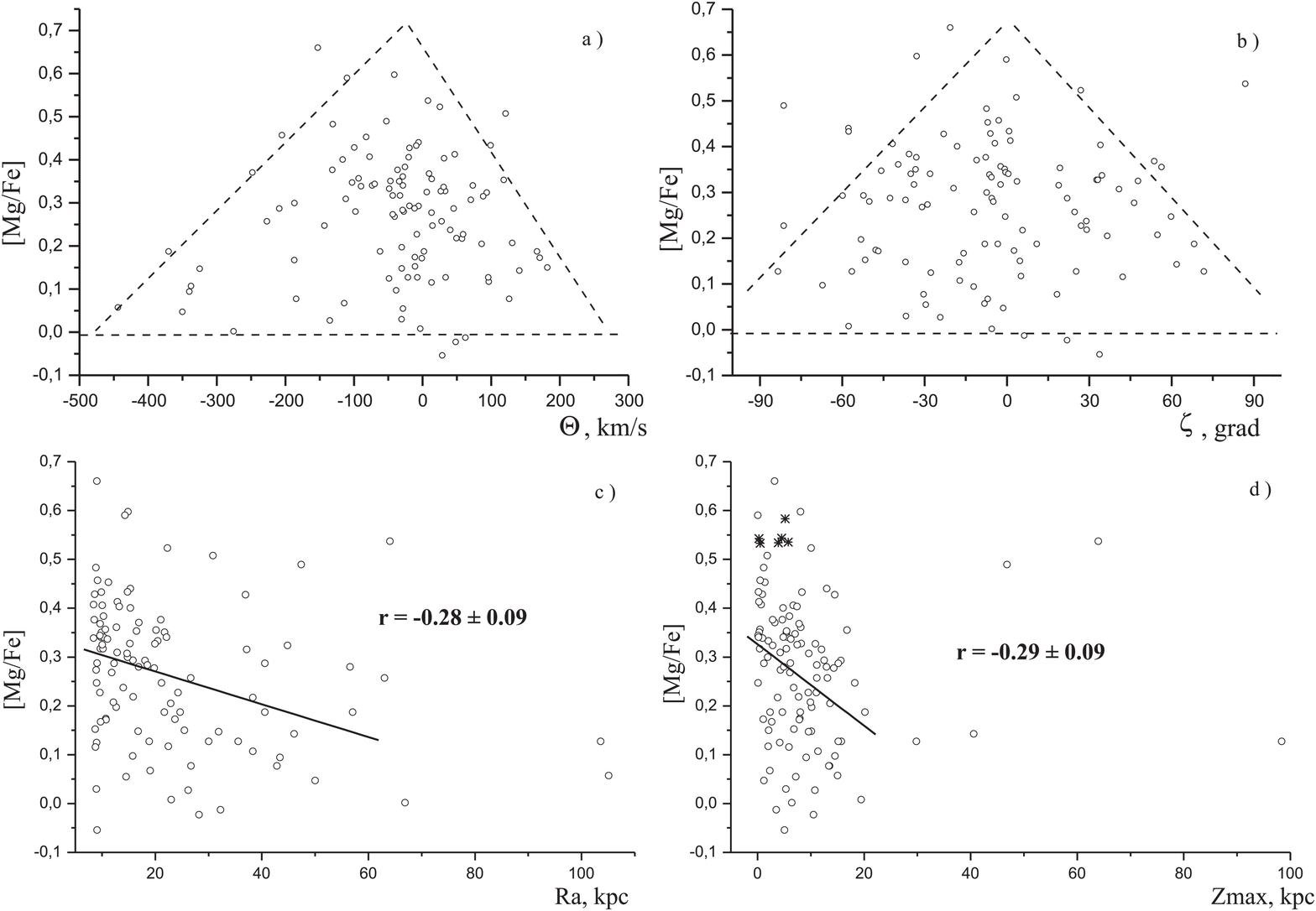}}
\caption{Relative magnesium abundances in accreted 
stars vs. their azimuthal velocities (a), Galactic 
orbital inclinations (b), maximum distances of the 
orbital points from the Galactic center (c) and plane (d). 
The dashed lines represent the envelopes of the 
points in the diagrams drawn by eye (upper row). The 
solid lines represent the regression lines 
for accreted halo stars (lower row).}
\label{fig_4}
\end{figure}

From the Fig.\,\ref{fig_4}\,a,\,b, where are substituted 
only accreted stars, one can see well, that (1) stars with 
the low azimuthal velocities and the small orbital 
inclinations are majority. (This is understandable, 
because comparatively massive satellite galaxies lose many 
stars.)  (2) Only star with small ($|\Theta |<50$~km\,s$^{-1}$) 
and with small orbit inclinations can have the high relative 
abundance of magnesium. (3) In contrast to them, the stars 
rapidly rotating around the Galactic center and stars with 
the large orbit inclinations, demonstrate in essence the low 
relations [Mg/Fe], uncharacteristic for such metal-poor stars. 
Further, (Fig\,\ref{fig_4}\,c,\,d) the negative radial and 
vertical gradients of the relative magnesium abundance also 
indicate the small relations [Mg/Fe] in the accreted stars 
with the extensive orbits. (These gradients reflect the 
sizes of the orbits, being located on which satellite 
galaxies lose their stars.)

Thus, sizes and inclinations of orbits in the accreted 
stars (and hence in their destroyed parent galaxies) 
increase with the decrease of the relative abundances of 
magnesium in them. The extensive and inclined orbits, 
according to the numerical simulation of the hierarchical 
formation of the galactic halo, as it was already said, 
one should expect in the debris of the low massive satellite 
galaxies, which are destroyed earlier than their orbit 
noticeably will change under the action of the tidal 
forces of the Galaxy.
Apparently, low massive galaxies, intersecting galactic 
plane, lose not only stars, but also interstellar gas while 
crossing the Galactic plane. Star formation in them ends 
fairly rapidly because of the loss of interstellar matter. 
Therefore in them we barely see any metal rich stars. In 
view of this the anomalously low [Mg/Fe] ratios in the 
lost by them metal-poor stars are caused by the not so 
much low star formation rate in their parental dwarf 
galaxies, as the fact that in the less massive dwarf 
galaxies the initial stellar mass function is just 
truncated at the high masses. As a result, SNe\,II eject 
into the interstellar medium a smaller amount of light 
$\alpha$-elements into the interstellar
medium and the [Mg/Fe] ratios for the stars become anomalously 
low compared the stars of the same metallicity that are 
genetically related to the Galaxy.

Thus, the properties of globular clusters and field 
stars discovered in the work are organically fit within 
the framework of a single hypothesis. According to it 
metal-poor stars with anomalously low $\alpha$-element 
abundances come into our Galaxy from debris of low-mass 
satellite galaxies in which the chemical evolution proceeded 
not only slowly but also with the absence of massive SNe\,II.

So, the results of comprehensive statistic studies testify 
that a significant quantity mainly of metal-poor objects, 
which belong at present to our Galaxy, were formed beyond 
its limits.  

{\it Acknowledgements.} This work was supported in part by the 
Federal Agency for Education (projects RNP~2.1.1.3483 and 
RNP~2.2.3.1.3950) and by the Southern Federal University 
(K07T~-- 125)\\[2.5mm]
\indent

{\bf References\\[2mm]}
1.~M.G.~Abadi, M.G.~Navarro, M.~Steinmetzand, and V.R.~Eke, Astrophys.J. 
{\bf 591}, 499, (2003).\\
2.~T.V.~Borkova and V.A.~Marsakov, Bull.Spec.Astrophys.Obs.
{\bf 54}, 61 (2002).\\
3.~T.V.~Borkova and V.A.~Marsakov, Pis'ma Astron.Zh.
{\bf 30}, 173 (2004) [Astron.Lett. {\bf 30}, 148 (2004)].\\
4.~V.A.~Marsakov and T.V.~Borkova, Pis'ma Astron.Zh. {\bf 32}, 545 (2006)
[Astron. Lett. {\bf 32}, 376 (2006)]\\
5.~T.~Tshuchiya, D.~Dinescu, and V.I.~Korchagin, Astrophys.J.
{\bf 589}, L29 (2003).\\

\vfill
%

\end{document}